\RequirePackage{fix-cm}
\documentclass[smallextended]{svjour3}       
\usepackage[italian,english]{babel}
\smartqed  
\usepackage{amsmath}
\usepackage{amssymb}
\usepackage{lipsum}
\usepackage{dsfont}
\usepackage{multirow}
\usepackage{afterpage}
\usepackage{amsfonts}
\usepackage{bm}%
\usepackage{babel}
\usepackage{hyperref}
\usepackage{graphicx}
\usepackage{color}
\usepackage{graphicx}
\begin{document}

\title{Dynamics of measurement induced nonlocality under decoherence 
}


\author{R. Muthuganesan        \and
       R. Sankaranarayanan 
}


\institute{Department of Physics, National Institute of Technology, Tiruchirappalli, India\at
              Tel.: +91-8760372825\\
              \email{rajendramuthu@gmail.com}           
           \and
           Department of Physics, National Institute of Technology, Tiruchirappalli, India \at
              \email{sankar@nitt.edu}
}

\date{Received: date / Accepted: date}

\maketitle

\begin{abstract}
Measurement Induced Nonlocality (MIN)-- captures nonlocal effects of a quantum state due to local von Neumann projective measurements, is a bona--fide measure of quantum correlation between constituents of a composite system. In this paper, we study the dynamical behavior of  entanglement (measured by concurrence), Hilber-Schmidt MIN and fidelity based MIN (F--MIN) under local noisy channels such as hybrid (consists of bit, phase, bit and phase flip), generalized amplitude damping (GAD) and depolarizing channels  for the initial Bell diagonal state. We observed that while sudden death of entanglement occur in hybrid and GAD channels,  MIN and F--MIN are more robust against such noise. Finally, we demonstrate the revival of MIN and F--MIN after a dark point of time against depolarizing noise.
\end{abstract}

\section{Introduction}
\label{intro}
Entanglement, a \textit{spooky--action} at distance exhibits in composite quantum system due to superposition principle, is originally introduced by Einstein \cite{Einstein1935} and Schr\"odinger \cite{Schrödinger1935}. Due to the pioneering work of Bell, entanglement is understood to be the manifestation of nonlocality which cannot be explained by local hidden variable theory \cite{Bell1964}. Nonlocality of a physical system, an unique feature of quantum mechanics having no classical analog, can be demonstrated through the violation of Bell inequality. Though entanglement is seen as a valuable resource for information processing, in certain context it can not be readily used. For example, bound entangled states are not useful in teleportation process \cite{Horodecki1999}. For tasks such as quantum cryptography \cite{Acin2007}, certified randomness generation \cite{Pironio2010} and quantum communication \cite{ Brukner2002}, the entangled states are useful only if they exhibit nonlocal correlations between the subsystems.

There exist other forms of quantum correlations even in the absence of entanglement, which is measured through quantum discord \cite{Ollivier2001}. This measure is shown to have advantage over the entanglement in deterministic quantum computation models using highly mixed  separable states and one qubit state \cite{Lanyon2008}. However, it has been observed that quantum discord is not calculable in closed form for an arbitrary two-qubit state \cite{Girolami2011}. Further, Luo and Fu defined the maximum nonlocal effect of locally invariant measurements as measurement induced nonlocality (MIN) \cite{SLUO2011}. This measure is in some sense dual to the geometric measure of quantum discord \cite{Dakic2010}. Similar to some of the other measures mentioned above, this measure quantifies the global effect of statistical mixture of states rather than the effect due to entanglement alone. Hence the above notion of nonlocality may also be considered to be more general than that of Bell's version of nonlocality.

In practice, any realistic system has an uncontrollable coupling to the environment which influences the time evolution or dynamical changes of the system. Dynamical evolution of open system is  described with an effective equation of motion - Master equation. Quantum operation is an alternative powerful technique for describing the evolution of a quantum system in various circumstances. It captures dynamical change of the system (phase change, depolarizing, decay of atom, etc.) due to the some physical process and the interaction between the system of interest and environment. Due to this coupling of system with environment the quantum correlation will suffer. Hence, it is important to study the dynamical behavior of quantum correlations when the system interacts with noisy environment. 

In this paper, we study the dynamical behavior of  entanglement (measured by the concurrence), MIN and fidelity based MIN  under local noisy channels such as hybrid (consists of bit, phase, bit and phase flip), generalized amplitude damping (GAD) and depolarizing channels by considering Bell diagonal state as an initial state. We extend our  analysis for both maximally and partially entangled states. We find that the environmental noise destroys the quantum correlations of the system. In particular, all the noise cause sudden death of entanglement (ESD) i.e, disentanglement occur in a finite time. Our investigations show that both MINs and fidelity based MIN decay  asymptotically, while entanglement suffer with sudden death. Further, we find an interesting result that both MINs revive after a dark point time when the system is subjected to depolarizing noise, while concurrence remains zero after ESD.

This paper is organized  as follows: In sec. \ref{correlation}, we briefly review the concept of entanglement, MIN and fidelity based MIN. In sec.\ref{Dynamics}, the dynamics of these measures are studied for the Bell diagonal state under local decoherence. Finally, the conclusions are presented in sec. \ref{Concl}. 

\section{Overview on correlation measures}
\label{correlation}

The amount of entanglement associated with a given two-qubit state $\rho$ can be quantified using concurrence \cite{Hill1997}, which is defined as 
\begin{equation}
C(\rho )=\text{max}\{0,~\lambda_1-\lambda_2-\lambda_3-\lambda_4\}
\end{equation}
where $\lambda_i$ are square root of eigenvalues of matrix $R=\rho \tilde{\rho }$ arranged in decreasing order. Here $\tilde{\rho } $ is spin flipped density matrix, which is defined as $\tilde{\rho }=(\sigma _y \otimes \sigma _y)\rho^{*}(\sigma _y \otimes \sigma _y)$. The symbol $*$ denotes complex conjugate in computational basis. It is known that $0\leqslant C(\rho ) \leqslant 1$ with minimum and maximum values correspond to separable and maximally entangled states respectively.

An arbitrary bipartite state $\rho $ in $\mathds{C}^2 \otimes \mathds{C}^2 $ can be represented as,
\begin{equation}
\rho= \sum_{i,j}\gamma _{ij}X_{i}\otimes  Y_{j} \label{state1}
\end{equation}
where $\Gamma = (\gamma _{ij} =\text{Tr} (\rho ~X_{i}\otimes  Y_{j}))$ is a $4 \times 4$ real matrix, $X_i$
and $Y_j$ are orthonormal operators for subsystem $a$ and $b$ respectively with $\text{Tr}(X_{k}X_{l})=\text{Tr}(Y_{k}Y_{l})=\delta _{kl}$. If $X_{0}=\mathds{1}^{a}/\sqrt{2}$, $Y_{0}=\mathds{1}^{b}/\sqrt{2}$, and separating the terms in eq.(\ref{state1}) the state $\rho$ can be written as  
\begin{equation}
\rho =\frac{\mathds{1}^{a} \otimes \mathds{1}^{b}}{4}+\sum_{i=1}^{3}x_{i}X_{i} \otimes \frac{\mathds{1} ^{b}}{\sqrt{2}}+\sum_{j=1}^{3}\frac{\mathds{1} ^{a}}{\sqrt{2}}\otimes y_{j}Y_{j}+\sum_{i,j\neq 0} t_{ij}X_{i}\otimes Y_{j}   \label{state2}
\end{equation}
 The components of Bloch vectors are $x_{i}=\text{Tr}[\rho (\sigma _{i} \otimes \mathds{1}^{b})]/2$ and $y_{i}=\text{Tr}[\rho (\mathds{1}^{a} \otimes \sigma _{j} )]/2$ with $t_{ij}=\text{Tr}[\rho (\sigma _{i}\otimes \sigma _{j})]/2$ being real matrix elements of correlation matrix $T$. 

Luo and Fu introduced a new measure to capture global effects of a quantum state due to local von Neumann projective measurements. It is originally defined as maximal square of Hilbert-Schmidt norm of difference of pre- and post- measurement state. Mathematically it is defined as \cite{SLUO2011}
\begin{equation}
 N(\rho ) =~^{\text{max}}_{\Pi ^{a}}\| \rho - \Pi ^{a}(\rho )\| ^{2} 
\end{equation}
where the maximum is taken over the von Neumann projective measurements on subsystem $a$. Here $\Pi^{a}(\rho) = \sum _{k} (\Pi ^{a}_{k} \otimes   \mathds{1} ^{b}) \rho (\Pi ^{a}_{k} \otimes    \mathds{1}^{b} )$, with $\Pi ^{a}= \{\Pi ^{a}_{k}\}= \{|k\rangle \langle k|\}$ being the projective measurements on the subsystem $a$, which do not change the marginal state $\rho^{a}$ locally i.e., $\Pi ^{a}(\rho^{a})=\rho ^{a}$. If $\rho^{a}$ is non-degenerate, then the maximization is not required and equal to the geometric discord \cite{Dakic2010}, which is the closest separation between the state under consideration and zero discord state.

In fact, it has closed formula for  $2\times n$ dimensional system  as 
\begin{equation}
N(\rho)=\begin{cases}
\text{Tr}(TT^t)-\frac{1}{\Vert \textbf{x} \Vert^2}\textbf{x}^tTT^t\textbf{x} & \quad \textbf{x}\neq 0, \\
\text{Tr}(TT^t)-\lambda_{\text{min}} & \quad \textbf{x}=0. \label{HS_MIN}
\end{cases}
\end{equation}
where $\lambda_{\text{min}}$ be the least eigenvalues of matrix $TT^t$. 

This quantity is easily computable and also experimentally realizable. However, it can change arbitrarily and reversibly through actions of the unmeasured party --local ancilla problem, indicated by Piani \cite{Piani2012}. Due to local ancilla problem MIN is not a bonafide measure to capture quantum correlation of a state. This issue can be resolved by replacing density matrix by its square root \cite{Chang2013}. Another natural way to circumvent this local ancilla problem is defining the correlation measure in terms of  sine metric defined as below.

Defining sine metric as $\mathcal{C}(\rho,\Pi^a(\rho))=\sqrt{1-\mathcal{F}(\rho, \sigma)}$ where $\mathcal{F}(\rho, \sigma)$ is the fidelity between the states $\rho$ and $\sigma $ defined as \cite{Wang2008}
\begin{equation}
 \mathcal{F}(\rho, \sigma) =\frac{(\text{Tr}(\rho\sigma))^2}{\text{Tr}(\rho)^2 \text{Tr}(\sigma)^2} \nonumber
\end{equation}
which statifies axioms of orginal fidelity \cite{Jozsa2015}. Multiplicative property of fidelity remedies the local ancilla problem correlation measure. Fidelity itself is not a metric, any monotonically decreasing function of fidelity defines a valid distace measure. Defining MIN based on fidelity induced metric as \cite{Muthu1,Muthu2}
\begin{equation}
N_{\mathcal{F}}(\rho ) = ~^{\text{max}}_{\Pi^a}~\mathcal{C}^2(\rho,\Pi^a(\rho)=~1-~^{\text{min}}_{\Pi^a}\mathcal{F}(\rho,\Pi^a(\rho))
\end{equation}
where minimization is carried over von Neumann projective measurements.

The closed formula of fidelity based MIN (F--MIN) for $2 \times n$ dimensional system is given as
\begin{equation}
N_{\mathcal{F}}(\rho)=  
\begin{cases}
\frac{1}{\| \Gamma  \|^{2}}(\| \Gamma  \|^{2}-\mu _{1}) & \text{if}~~~ \textbf{x}=0 \\ \frac{1}{\| \Gamma  \|^{2}}(\| \Gamma  \|^{2}-\epsilon ) & \text{if}~~~\textbf{x}\neq0
\end{cases}
\end{equation}
where $\epsilon =\text{Tr}(A\Gamma \Gamma ^{t}A^{t})$, $\mu _{1}$ is the minimum eigenvalue of matrix $\Gamma \Gamma ^{t}$ and 
\begin{equation} \label{eq:ope}
A=\frac{1}{\sqrt{2}}
\begin{pmatrix}
1 & \frac{\textbf{x}}{\| \textbf{x} \|}\\
1 & -\frac{\textbf{x}}{\| \textbf{x} \|}
\end{pmatrix} \nonumber
\end{equation} 
with $a_{ki}=\langle k| X_{i}|k \rangle $ $(i=0,1,2,3)$.
\section{MIN under decoherence}
\label{Dynamics}

In this section, we study the dynamics of quantum correlation measured by  MIN and F--MIN and compare with entanglement (measured by concurrence) under local decoherence channels such as hybrid, depolarizing and generalized amplitude damping channels. The initial state for our investigation is Bell diagonal state whose marginal state is maximally mixed. The Bloch representation of the state is given by 
\begin{equation}
\rho^{BD}=\frac{1}{4}\left[\mathds{1}\otimes\mathds{1}+\sum^3_{i=1}c_i(\sigma_i \otimes \sigma_i)\right]
\end{equation}
where $\mathds{1}$ is identity matrix, $\sigma_i$ are Pauli spin matrices and vector $\vec{c}=(c_1,c_2,c_3) $ is a three dimensional vector composed of correlation coefficients such that $-1\leq c_i=\langle \sigma_i \otimes\sigma_i \rangle \leq 1$ completely specify the quantum state. In matrix form, we have
\begin{equation}
\rho^{BD}=\frac{1}{4}
\begin{pmatrix}
1+c_3 & 0 & 0 & c_1-c_2 \\
0 & 1-c_3 & c_1+c_2 & 0  \\
0 & c_1+c_2 & 1-c_3 &0  \\
c_1-c_2 & 0 & 0 & 1+c_3
\end{pmatrix}.
\end{equation}
As the name implies, Bell diagonal state has four maximally entangled Bell state as eigenvectors, with corresponding eigenvalues 
\begin{equation}
\mu_{i,j}=\frac{1}{4}\left[1+(-1)^i c_1-(-1)^{i+j}c_2+(-1)^j c_3\right].
\end{equation}

The concurrence of the state is given by 
\begin{equation}
C(\rho^{BD})=2 ~\text{max} \{ 0, |c_1-c_2|  -(1-c_3), | c_1+c_2|  -(1+c_3) \}. \label{Conc}
\end{equation}
Then MIN and F-MIN of Bell diagonal state are computed as 
\begin{eqnarray}
N(\rho^{BD})=\frac{1}{4}\left(\sum_{i=1}^{i=3}c_i^2-c_0^2\right),  \nonumber \\ 
N_\mathcal{F}(\rho^{BD})=\frac{1}{\|\Gamma  \|^{2} }\left(\sum_{i=1}^{i=3}c_i^2-c_0^2\right), \label{MINBELL}
\end{eqnarray}
where $c_0=\text{min}\{ \lvert c_1 \rvert, \lvert c_2 \rvert, \lvert c_3 \rvert \} $. If $\rho^{BD}$ describes a valid physical state, then $0\leq \mu_{i,j}\leq 1$ and $\sum_{i,j}\mu_{i,j}=1$. Under this constraint, the vector $\vec{c}=(c_1, c_2, c_3) $ must be restricted to the tetrahedron whose vertices  are $(1, 1, -1)$, $(-1, -1, -1)$, $(1, -1, 1)$ and $(-1, 1, 1)$ \cite{Sarandy2013}. The vertices are easily identified as Bell states (EPR pairs), for which MIN and F--MIN are maximum, i.e., $N(\rho^{BD})=N_\mathcal{F}(\rho^{BD})=0.5$. In fact, since $\|\Gamma  \|^{2}=(1+c_1^2+c_2^2+c_3^2)/4\leq 1$ and we have the inequality $N(\rho^{BD})\leq N_\mathcal{F}(\rho^{BD})$, with equality holds for maximally entangled states.  Further, both the MIN and F-MIN are vanishing for the correlation vector $\vec{c}=(0, 0, 0) $, at which the state $\rho^{BD}=\mathds{1}/4$ is maximally mixed state.

The interaction of above quantum state with environment can be conveniently investigated using operator--sum representation. In this formalism, the evolution of quantum state is described by positive and trace preserving operation \cite{Nielsen2010}

\begin{equation}
\mathcal{E}(\rho)=\sum_{i,j}(E_i\otimes E_j) \rho (E_i\otimes E_j)^{\dagger} \label{Qoperation}
\end{equation}
where $\{ E_k\} $ is a set of Kraus operators associated with decohering process of a single qubit and satisfy the completeness property $\sum_kE_kE^{\dagger}_k=\mathds{1}$. We shall note that Bell diagonal state preserves its structure after the intervention of environment as described above .  Time evolved state is then given by
\begin{equation}
\mathcal{E}(\rho^{BD})=\frac{1}{4}
\begin{pmatrix}
1+c'_3 & 0 & 0 & c'_1-c'_2 \\
0 & 1-c'_3 & c'_1+c'_2 & 0  \\
0 & c'_1+c'_2 & 1-c'_3 &0  \\
c'_1-c'_2 & 0 & 0 & 1+c'_3
\end{pmatrix}
\end{equation}
with the correlation vector $\vec{c'}=(c'_1, c'_2, c'_3)$. Here the primed components are time dependent and unprimed are initial conditions. 
\begin{table}
\caption{Kraus operators for the bit flip (BF), phase flip (PF), bit--phase flip (BPF) \cite{Nielsen2010}}
\label{tab:1}       
\begin{tabular}{lll}
\hline\noalign{\smallskip}
Channel & ~~~$E_0$& ~~~~~~~~$E_1$  \\
\noalign{\smallskip}\hline\noalign{\smallskip}
BF & $\sqrt{1-p/2}\mathds{1}$ & $\sqrt{p/2} \sigma_1$ \\
PF & $\sqrt{1-p/2}\mathds{1}$ & $\sqrt{p/2} \sigma_3$ \\
BPF & $\sqrt{1-p/2}\mathds{1}$ & $\sqrt{p/2} \sigma_2$ \\
\noalign{\smallskip}\hline
\end{tabular}
\end{table}
\begin{figure*}[!ht]
\centering\includegraphics[width=0.8\linewidth]{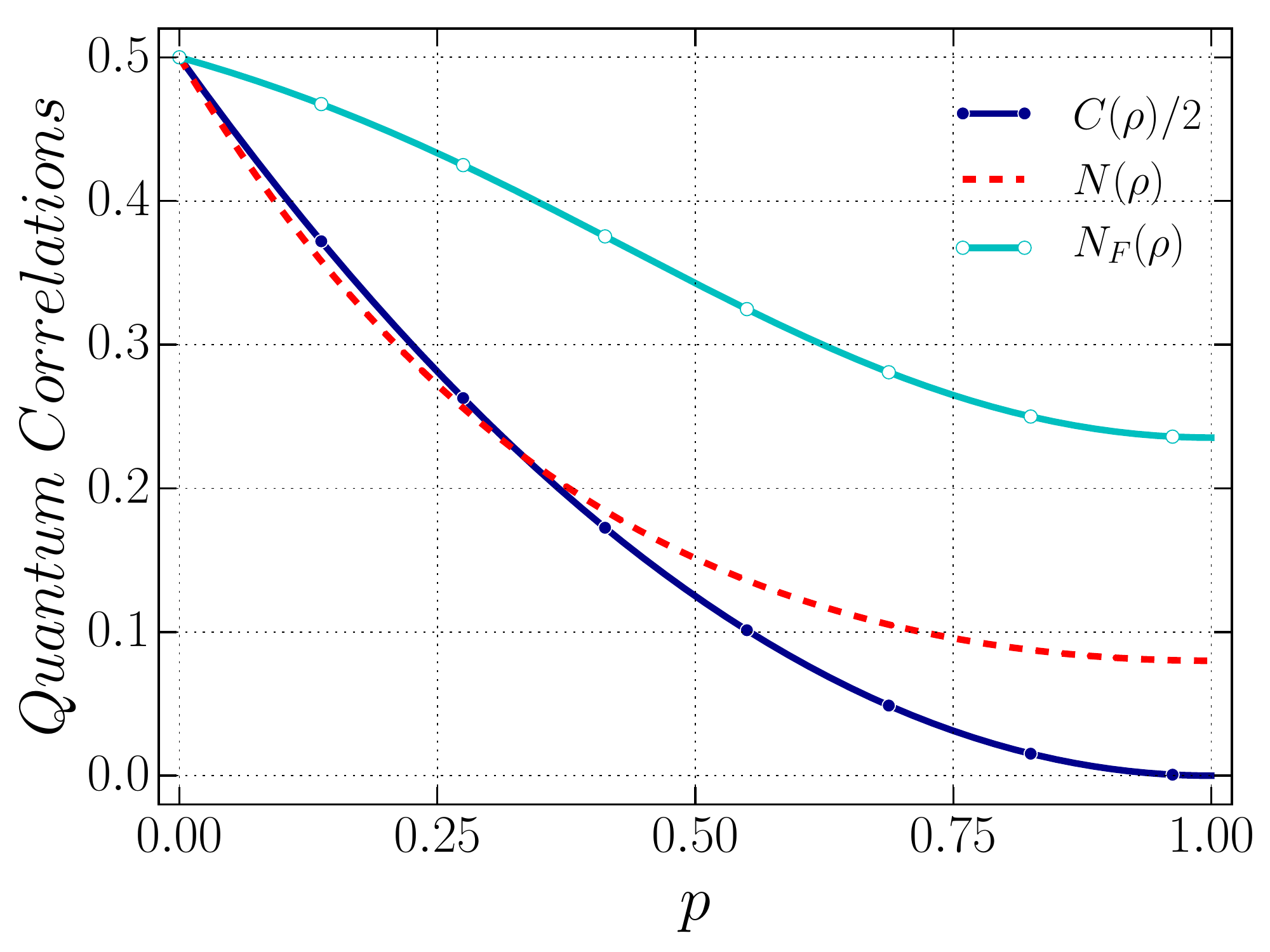}
\centering\includegraphics[width=0.8\linewidth]{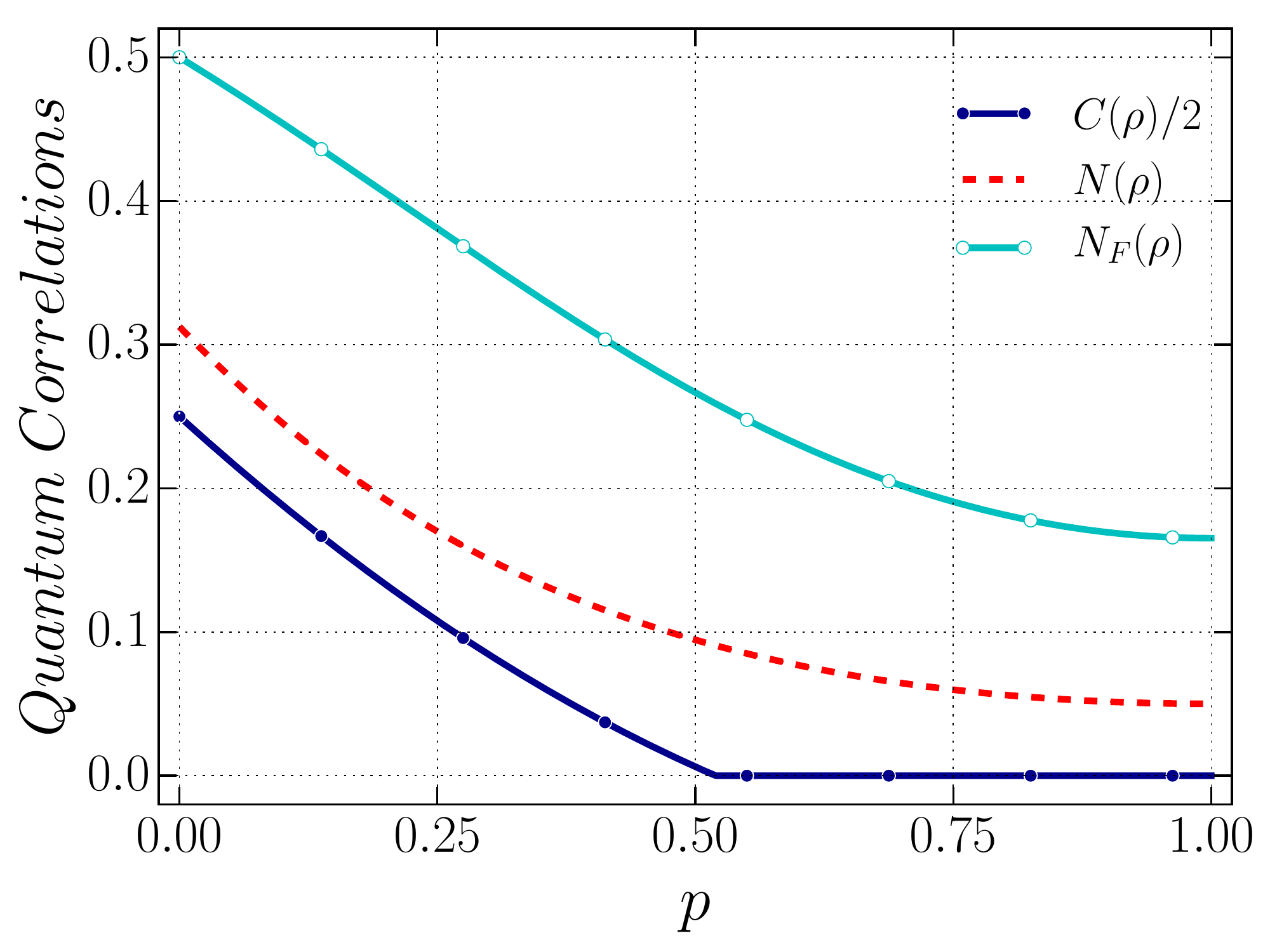}
\caption{(color online) Dynamical behaviour of quantum correlations under hybrid channel with $\alpha=\beta=0.4 ~\text{and} ~  \gamma=0.2$ for the initial  pure maximally entangled state with  $\vec{c}=(1,1,-1)$ (top) and mixed partially entangled state $\vec{c}=(1,0.5,-0.5)$ (bottom).}
\label{fig2}
\end{figure*}

(i) \textit{Hybrid channel}: Now, we consider a qubit subjected to multi--channel with unequal probability i.e., hybrid channel. Here the hybrid channel consists of bit flip, bit-phase flip,  phase flip with probability of $\alpha$, $\beta$ and $\gamma$ respectively. Defining $p~(0\leq p \leq 1)$ as probability the respective Kraus operators are as given in table \ref{tab:1}. Hence the time evolved state is given as
\begin{equation}
\mathcal{E}(\rho^{BD})=\alpha \mathcal{E}_{BF}(\rho^{BD})+ \beta \mathcal{E}_{BPF}(\rho^{BD})+\gamma \mathcal{E}_{PF}(\rho^{BD})
\end{equation}
with $\alpha+\beta+\gamma=1$. The components of  correlation vector of time evolved state are given as 
\begin{eqnarray}
c'_1=\left[\alpha+(\beta+\gamma)(1-p)^2\right]c_1,  \nonumber \\ 
c'_2=\left[\beta+( \alpha+\gamma)(1-p)^2\right]c_2,  \nonumber \\
c'_3=\left[\gamma +(\alpha+\beta)(1-p)^2\right]c_3.
\end{eqnarray}

In what follows, we first study the dynamics of MIN and F--MIN along with entanglement (measured by concurrence) for different initial conditions under hybrid channel. In Fig \ref{fig2}, we plot the dynamical behavior of MIN, F--MIN and concurrence as function of decoherence parameter $p$ in hybrid channel with $\alpha=\beta\neq \gamma$. Taking $\alpha=\beta=0.4$,  $\gamma=0.2$ and initial state with $\vec{c}=(1, 1, -1) $, which is a pure and maximally entangled state $(| 01 \rangle+| 10 \rangle)/\sqrt{2}$, one of the vertices of tetrahedron of Bell diagonal state as mentioned earlier \cite{Sarandy2013}.  From  Fig.\ref{fig2}, one can find that all the correlation measures decreases with parameter $p$ and concurrence vanishes at $p=1$. On the other hand, the companion quantities are not zero anywhere, which shows the presence of correlation between the subsystems even in the absence of entanglement. 

Next, we consider a partially entangled mixed state as initial state with coordinates $\vec{c}=(1, 0.5, -0.5) $. Unlike the dynamics of pure maximally entangled state, in this case the hybrid channel causes early death in concurrence with respect to decoherence parameter $p$. In other words, the concurrence vanishes for $p\geq  (p_c)_\pm$ where  
\begin{equation}
 (p_c)_\pm=1-\left[\frac{1-\alpha c_\pm \pm c_3 \gamma}{(\alpha+ \gamma)c_\pm \mp 2 \alpha c_3}\right]^{1/2}
\end{equation}
with $c_\pm=\lvert c_1\pm c_2\rvert $. This phenomenon is called the sudden death of entanglement with respect to the parameter \cite{Eberly2009}. For the above examples $\vec{c}=(1, 1, -1) $ and $\vec{c}=(1, 0.5, -0.5) $  we have $p_c=1$ and $p_c=0.53$ respectively. Since MIN and F--MIN  do not vanish for any values of $p$,  the results presented above suggest that MIN and F--MIN are more robust against decoherence in capturing correlation in the given system. 
\begin{figure*}[!ht]
\centering\includegraphics[width=0.8\linewidth]{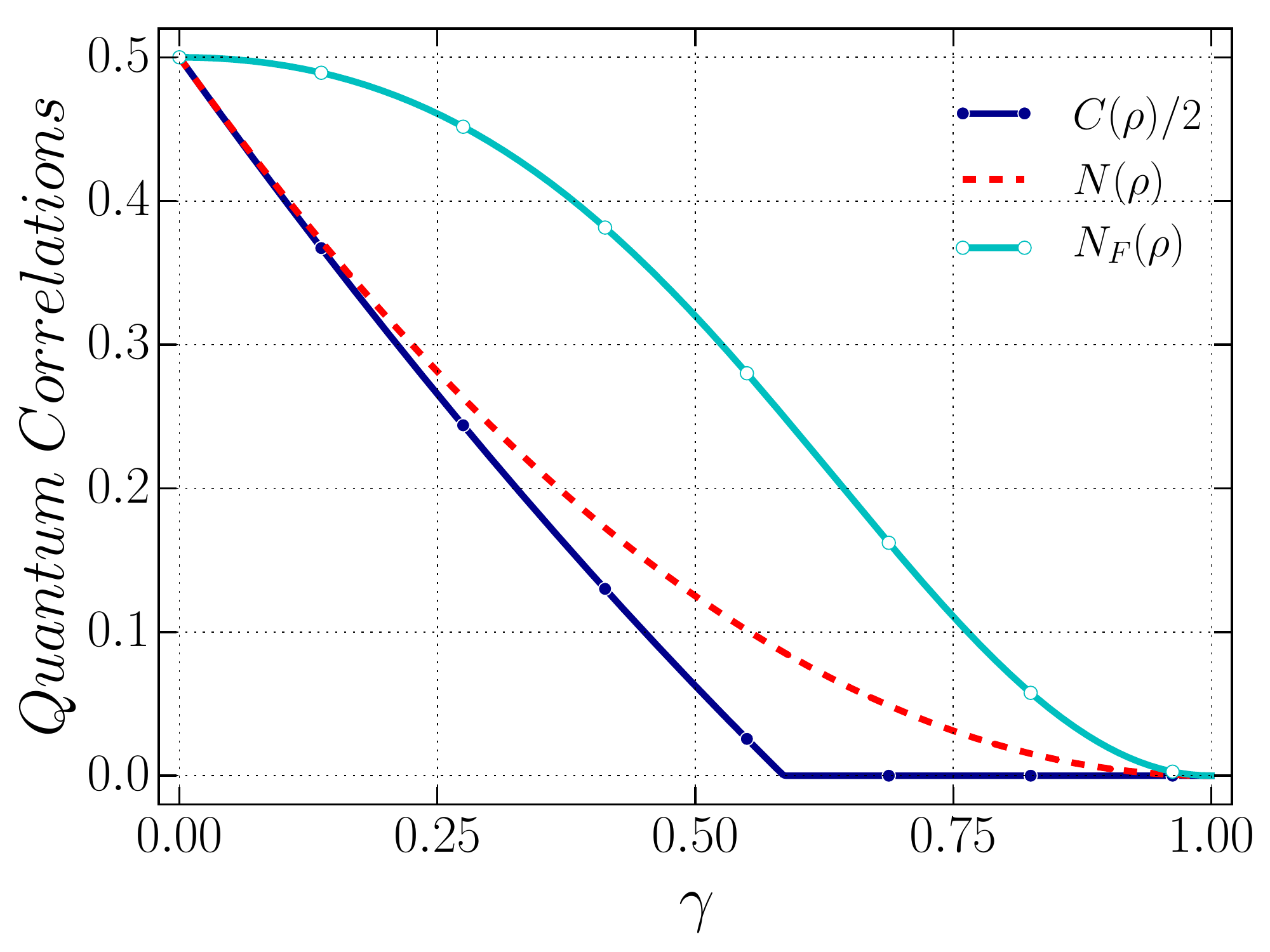}
\centering\includegraphics[width=0.8\linewidth]{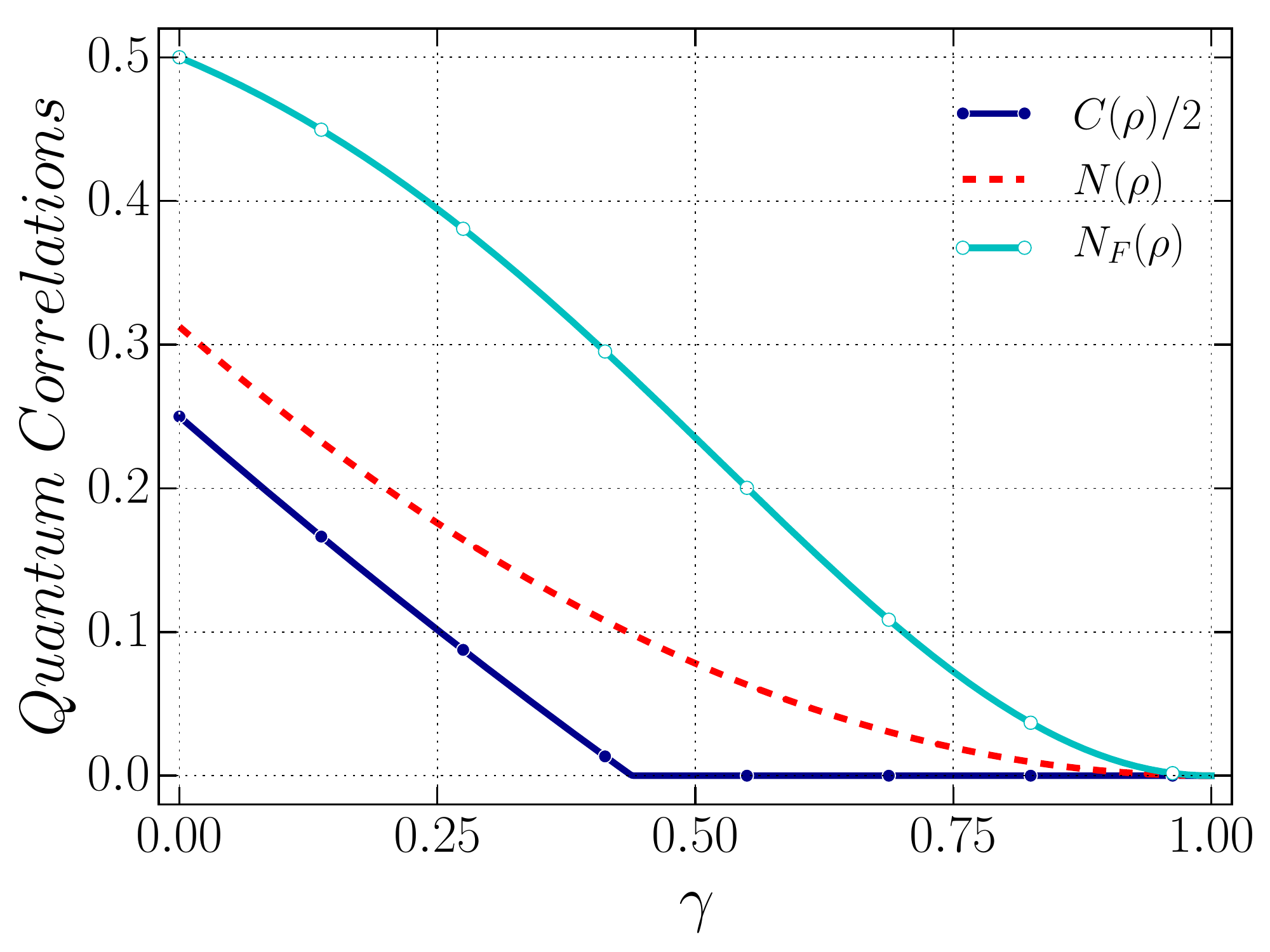}
\caption{(color online) Dynamical behavior of quantum correlations under generalized amplitude damping channel for the initial pure maximally entangled state with  $\vec{c}=(1,1,-1)$ (top) and mixed partially entangled state $\vec{c}=(1,0.5,-0.5)$ (bottom).}
\label{GAD}
\end{figure*}

(ii) \textit{Generalized Amplitude Damping:}
Here we consider the generalized amplitude damping (GAD), which models the loss of energy from quantum system to environment at a finite temperature such as thermal bath. Such a process is described by the Kraus operators \cite{Nielsen2010}
\begin{eqnarray}
E_{0}&=&\sqrt{p}
\begin{pmatrix}
1 & 0\\
0 & \sqrt{1-\gamma }
\end{pmatrix},
~E_{1}=\sqrt{p}
\begin{pmatrix}
0 & \sqrt{\gamma }\\
0 & 0
\end{pmatrix}, \nonumber \\
E_{2}&=&\sqrt{1-p}
\begin{pmatrix}
\sqrt{1-\gamma} & 0\\
0 &  1
\end{pmatrix},
~E_{3}=\sqrt{1-p}
\begin{pmatrix}
0 & 0\\
 \sqrt{\gamma } & 0
\end{pmatrix}, \nonumber
\end{eqnarray} 
where $\gamma =1-\mathrm{e}^{-\gamma' t}$, $\gamma' $ is decay rate and $p$ defines the final probability distribution of stationary (equilibrium) state. For simplicity we fix $p=1/2$ and the components of evolved state under this channel are given by
\begin{equation}
  c'_1=(1-\gamma)c_1,~~~ c'_2=(1-\gamma)c_2, ~~~~c'_3=(1-\gamma)^2 c_3.
\end{equation}
Here also we study the dynamical behaviour of correlation measures for same initial conditions as given earlier. 

Firstly, we consider the pure maximally entangled state with the  initial condition $\vec{c}=(1, 1, -1) $. On substituting the evolved state elements in the expression of concurrence, we find that entanglement of the evolved state is zero for $\gamma\geq \gamma_0\simeq 0.58$ as shown in Fig. \ref{GAD}. In other words, evolution of the maximally entangled state under this channel exhibits entanglement sudden death i.e., influence of quantum noise reduces the entanglement to zero in finite time. It is also clear from our results that, dynamics of both the MINs are qualitatively same. In particular, as time increases the MIN and F-MIN decrease showing that quantum correlation vanishes asymptotically. 

One can observe similar effect for the partially entangled state with vector $\vec{c}=(1, 0.5, -0.5) $ and the entanglement of the evolved state is zero for $\gamma\geq \gamma_0\simeq 0.4$. In this channel also the non-zero MINs in the region of zero concurrence show the existence of quantum correlation without entanglement. Hence, we conclude that MIN and F-MIN are more robust than the entanglement measure against GAD channel.

\begin{figure*}[!ht]
\centering\includegraphics[width=0.8\linewidth]{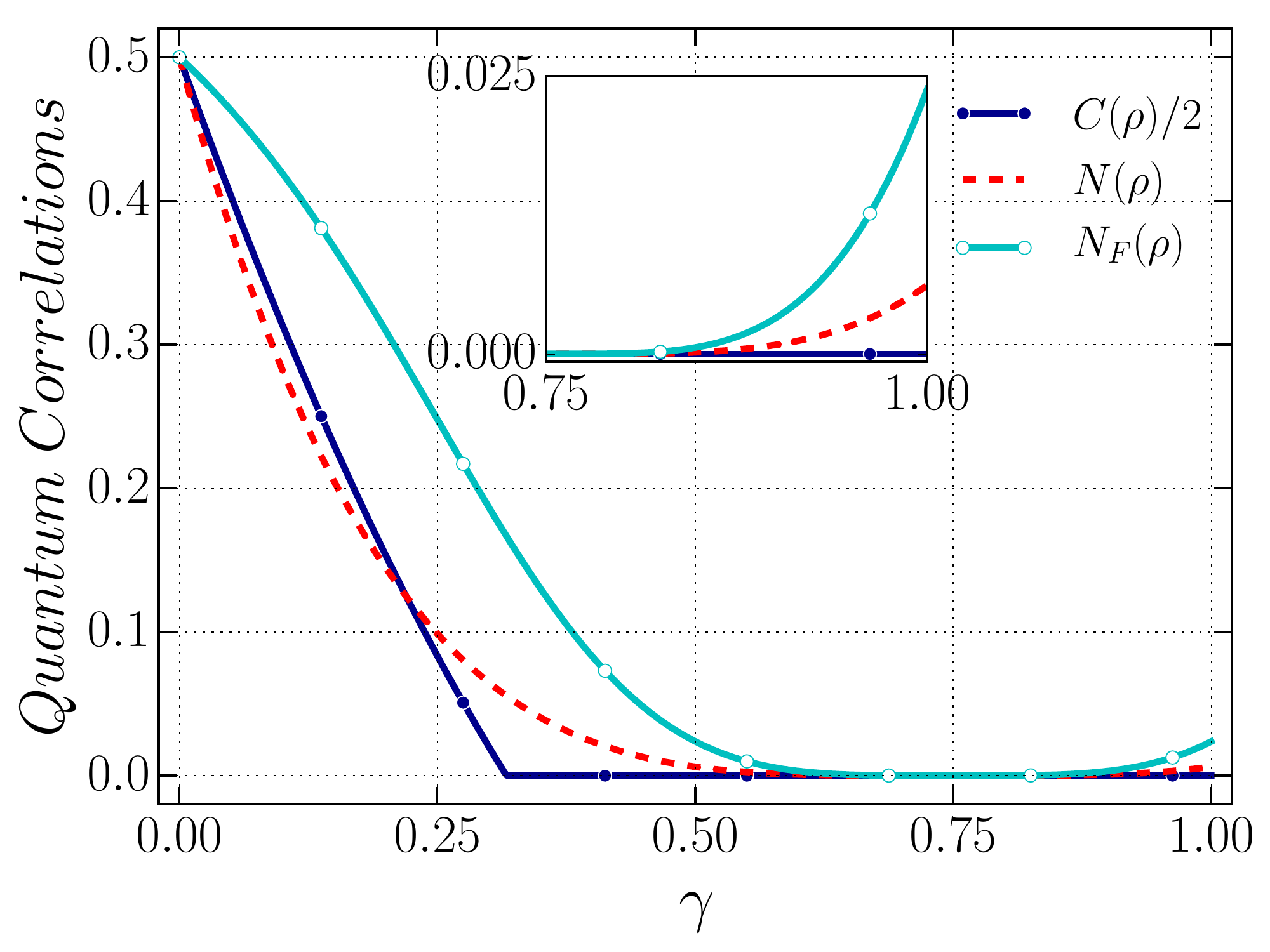}
\centering\includegraphics[width=0.8\linewidth]{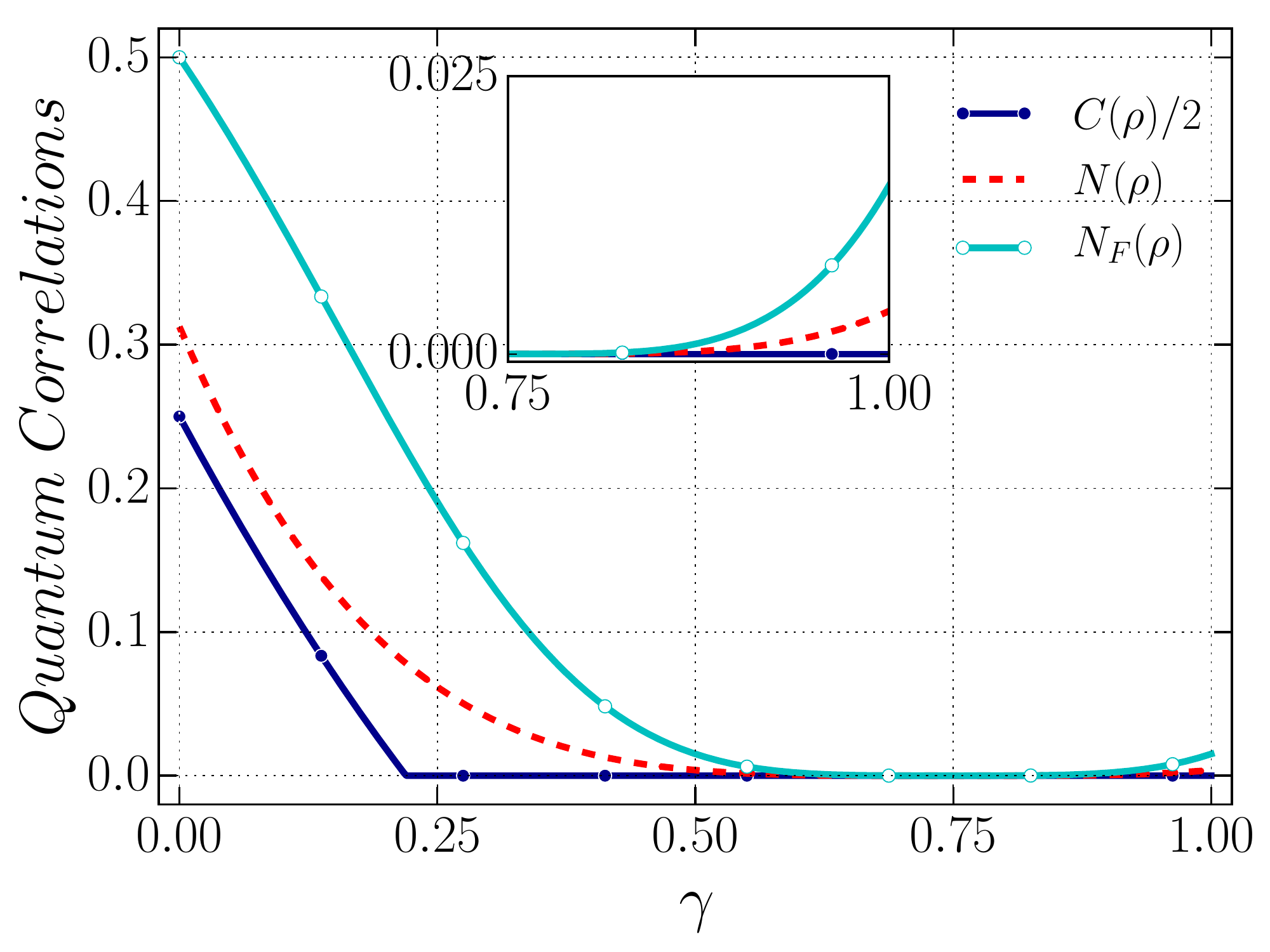}
\caption{(color online) Dynamical behavior of quantum correlations under depolarizing channel for the initial pure maximally entangled state with  $\vec{c}=(1,1,-1)$ (top) and mixed partially entangled state $\vec{c}=(1,0.5,-0.5)$(bottom).}
\label{depol}
\end{figure*}

(iii) \textit{Depolarizing channel}:
This channel is a type of quantum noise which transforms a single qubit into a maximally mixed state $\mathds{1}/2$ with probability $\gamma $ and   $1-\gamma$ is the  probability that the qubit remains intact. This channel is represented by the Kraus operators \cite{Nielsen2010}: 
\begin{eqnarray}
E_{0}&=&\sqrt{1-\gamma} ~\mathds{1}, ~E_{1}=\sqrt{\gamma/3} ~\sigma _{1} \nonumber\\ 
E_{2}&=&\sqrt{\gamma/3}~ \sigma _{2},~E_{3}=\sqrt{\gamma/3}~ \sigma _{3} \nonumber
\end{eqnarray}
where  $\gamma =1-\mathrm{e}^{-\gamma't}$ with $\gamma'$ being damping constant. The time evolved state coefficients are 
\begin{equation}
c'_i=\left(\frac{4\gamma}{3}-1\right)^2 c_i ~~~~~~~~~\text{with}~~ i=1,2,3. 
\end{equation}
Up on substituting the density matrix elements in Eqs. (\ref{Conc}) and (\ref{MINBELL}) we obtain concurrence, MIN and F-MIN for this channel. Form Fig. \ref{depol} we obtain the sudden death of entanglement for the initial states. The MIN and F--MIN are also shown to decrease with time.  On the other hand, unlike the hybrid and GAD channels both the MIN and F-MIN vanishes for a  passage of time. Remarkably, the MIN and F-MIN revive back after that dark point (zero MINs) of time. 

\section{Conclusions}
\label{Concl}
We have studied the dynamical behaviour of  entanglement (measured by the concurrence), Hilbert-Schmidt and fidelity based MINs under various noisy channels by considering Bell diagonal state as initial state. When the system is in maximally entangled state and subjected to hybrid channel (consists of bit, phase, bit and phase flip), the entanglement is vanishes asymptotically. For partially entangled state the hybrid channel cause sudden death of entanglement and other companion quantities are non-zero even in the asymptotic limit. In GAD channel, we show that both MIN and F-MIN vanish asymptotically but the noise causes sudden death of entanglement even for maximally entangled state. Unlike the earlier case, MIN, F--MIN and concurrence are vanishing for a passage of time against depolarizing noise. The MINs revives  after a dark point and while concurrence remains zero. In light of the above results, we have shown with sufficient evidence that MIN and F--MIN are robust against decoherence than the entanglement.

%
%




\end{document}